\documentclass[fleqn,usenatbib]{mnras}

\usepackage{newtxtext,newtxmath}
\usepackage{comment}

\usepackage[T1]{fontenc}

\DeclareRobustCommand{\VAN}[3]{#2}
\let\VANthebibliography\thebibliography
\def\thebibliography{\DeclareRobustCommand{\VAN}[3]{##3}\VANthebibliography}

\usepackage{graphicx}	
\usepackage{amsmath}	

\newcommand{\gp}{\mathrm{g^\prime}} 
\newcommand{\rp}{\mathrm{r^\prime}} 
\newcommand{\ip}{\mathrm{i^\prime}} 
\newcommand{\cps}{\mathrm{photons/s/cm^2}} 
\newcommand{\diff}[1]{\mathrm{d}#1}
\newcommand{\rout}{r_\mathrm{out}}
\newcommand{\LX}{L_\mathrm{X}}
\newcommand{\Cirr}{\mathcal{C}} 

\title[
    Optical/X-ray variations in Aql X-1 outbursts
]{
    Optical and X-ray variations during 5 outbursts of Aql X-1 in 3.6 years from 2016
}

\author[M. Niwano et al.]{
    Masafumi Niwano,$^{1}$\thanks{E-mail: niwano@hp.phys.titech.ac.jp}
    Katsuhiro L. Murata,$^{2}$
    Naohiro Ito,$^{1}$
    Yoichi Yatsu,$^{1}$
    and Nobuyuki Kawai$^{1}$
    \\
    $^{1}$ Department of Physics, Tokyo Institute of Technology, 2-12-1 Ookayama, Meguro-ku, Tokyo 152-8551, Japan\\
    $^{2}$ Okayama Observatory, Kyoto University, 3037-5 Honjo, Kamogata-cho, Asakuchi, Okayama 719-0232, Japan\\
}

\date{Accepted XXX. Received YYY; in original form ZZZ}

\pubyear{2023}

\begin{document}
\label{firstpage}
\pagerange{\pageref{firstpage}--\pageref{lastpage}}
\maketitle

\begin{abstract}
We analyzed optical/X-ray quasi-simultaneous light curves of Aql X-1, obtained by MAXI (Monitor of All-sky X-ray Image), ZTF (Zwicky Transient Facility) and LCO (Las Cumbres Observatory) in about 3.6 years from 2016, for understanding electromagnetic radiation mechanisms during its outbursts.
As a result, we confirmed that 5 outbursts had detected in the epoch, and that 3 outbursts underwent the X-ray state transition across Low-Hard, In-Transition, and High-Soft state while remaining 2 outbursts stayed in the Low-Hard state.
We found that the optical spectral energy distribution in the High-Soft state is consistent with a simplified irradiated disk model, and that the optical color/magnitude variation can be explained by variations in the X-ray luminosity and the disk geometrical thickness.
\end{abstract}




\section{Introduction}\label{sec:intro}
A Low-mass X-ray binary (LMXB) consists of a compact main star and a low mass ($\lesssim 1 M_{\sun}$) donor star.
In LMXB systems, masses accreting from the donor to the compact star by Roche-lobe overflow form structures such as an accretion disk, which emit electromagnetic radiation in a wide range of wavelengths from radio to X-ray (sometimes gamma-ray).
In particular, X-ray and optical radiations are emitted mainly from the vicinity of the compact star and the outer rim of the accretion disk, respectively.
Therefore, optical and X-ray simultaneous observations of LMXBs are important for studying the dynamics across the entirety of X-ray binary system.
Some LMXBs exhibit an outburst, a sudden and explosive brightening phenomena, and such LMXBs are called as Soft X-ray Transients (SXTs).
LMXBs have similar properties to cataclysmic variables (CVs) and their outbursts are interpreted by the disk instability model (DIM; \citealt{lasota,hameury}).
However, the DIM cannot fully explain behaviors of LMXB outbursts, and it is believed that X-ray irradiation to the disk plays an important role in LMXBs \citep{paradijs,dubus1999}.

Aql X-1 is a SXT composed of a neutron star \citep{koyama} and a K-type donor \citep{chevalier,sanchez}, and is known to undergo recurrent outbursts about once a year which are detected in the radio to X-ray band \citep{kaluzienski,priedhorsky,maitra_2008}.
In addition, hard-to-soft and soft-to-hard transitions are observed during brightening and decay phase of outbursts, respectively \citep{maitra2004,yu}.
There is an interloper star $\sim0\arcsec.5$ east of Aql X-1 \citep{chevalier,sanchez}, and it is difficult to resolve these 2 objects with ground-based optical observations without adaptive optics.
Furthermore, the interloper is $\sim$ 2 mag brighter than the quiescent magnitude of Aql X-1 in V-band.
Therefore, it is essential to remove the effect of interloper for analyzing the ground-based optical data of Aql X-1.

Since its discovery in 1967 \citep{friedman}, Aql X-1 has been the subject of numerous studies by predecessors.
\citet{maitra_2008} showed that the state of Aql X-1 in outburst can be classified into FRED (Fast Rise and Exponential Decay) and LIS (Low Intensity State) according to the light curve shape, and optical-infrared (OIR) variations during outburst can be explained by temperature changes of the irradiated disk outer rim, based on OIR and X-ray observations for 10 outbursts which occurred in 10 years from 1998.
\citet{tudose} studied the disk-jet coupling in Aql X-1 using radio, optical and X-ray data in 1986-2008.
As a result, they showed that the disk-jet coupling in Aql X-1 is similar to what has been observed in black hole X-ray binaries (BHXBs), and suggested that neutron star X-ray binaries (NSXBs) can mimic the behaviour of BHXBs in suppressing the jet in soft/disk-dominated X-ray states.
\citet{meshcheryakov} analyzed near-infrared (NIR), optical, near-ultraviolet (NUV) and X-ray data of the outburst June 2013, and revealed that NIR and NUV spectra can be explained by the disk irradiated with direct soft X-rays ($<10$ keV) and scattered hard X-rays ($>10$ keV) respectively, before and immediately after the state transition, and at the peak.
\citet{trigo} confirmed the change of the jet break frequency at the hard to soft transition and during decay in the 2016 outburst using radio, optical and X-ray data, and suggested that the properties of the internal jet rely most critically on the accretion disk and corona, rather than the compact object itself.
\citet{navas} investigated the UV/optical and X-ray correlation for outbursts in 2013, 2014 and 2016, and suggested that multiple emission processes contributed to the UV/optical emission of Aql X-1, such as the viscous heated disk or a hot flow.

The purpose of this research is further understanding of the electromagnetic radiation mechanism of Aql X-1 during its outbursts.
Therefore, we analyzed optical and X-ray light curves of Aql X-1 using 3.6 years of public data from MAXI, ZTF and LCO, and interpret optical and X-ray variations in the outbursts using a simplified irradiated disk model.
This paper organized as follows.
Information of observations and data reductions are described in section~\ref{sec:obs}.
Section~\ref{sec:result} summarized the results of data analysis.
In section~\ref{sec:discussion}, we interpret observational results using the simplified irradiated disk model.
Finally, section~\ref{sec:summary} summarizes the achievements and concludes this paper.

\section{Observations and data reduction}\label{sec:obs}
We used X-ray data of MAXI and optical data of ZTF and LCO.
In this paper, optical magnitudes are presented in AB system.
We considered a data-set taken with a time-difference of $\leq$ 0.5 days to be a simultaneous observation.
The error shall indicate $1\sigma$ error unless otherwise noted.
\autoref{tab:filters} describes typical wavelengths and frequencies of ZTF and LCO filters. 

\begin{table}
    \caption{Optical filters of ZTF and LCO data}
    \begin{tabular}{cccc}\hline
        Filter & Typical wavelength & Typical frequency & Reference\\
         & (\AA) & ($10^{14}$Hz) &\\\hline
        ZTF-$\gp$ & 4722 & 6.4 & *\\
        LCO-V & 5448 & 5.5 & $\dagger$\\
        ZTF-$\rp$ & 6339 & 4.7 & *\\
        LCO-$\ip$ & 7503 & 4.0 & $\dagger$\\\hline
        \multicolumn{4}{l}{
            * \url{http://svo2.cab.inta-csic.es/svo/theory/fps3/index.php}
        }\\
        \multicolumn{4}{l}{
            $\dagger$ \url{https://lco.global/observatory/instruments/filters/}
        }
    \end{tabular}
    \label{tab:filters}
\end{table}

\subsection{MAXI}\label{sec:maxi}
Monitor of All-sky X-ray Image \citep{maxi} is an X-ray camera installed onboard
the International Space Station.
We obtained two types of Gas Slit Camera (GSC) data via MAXI on-demand web interface\footnote{\url{http://maxi.riken.jp/mxondem/}}.
The first data-set comprises daily-averaged 2-6 keV and 6-20 keV light curves covering the Modified Julian Date (MJD) range 57500-58800.
These energy ranges are the default values of the MAXI on-demand.
The second consists of daily-averaged spectra acquired in the dates when simultaneous optical data was available.

We estimated the daily-averaged 2-20 keV energy-flux by fitting the spectrum using \texttt{xspec} \citep{xspec}.
We used \texttt{phabs*diskbb} model and successfully fitted with $\chi_r^2\approx 1$.

\subsection{ZTF}\label{sec:ztf}
Zwicky Transient Facility \citep{ztf} is an optical wide-field survey project
using a CCD camera array attached to the Samuel Oschin Telescope
at the Palomar Observatory in California.
We used $\gp$ and $\rp$-band lightcurves acquired from ZTF Public Data Release 3 (DR3; \citealt{zsds}).
DR3 covers the epoch of 58194-58848 MJD.
The typical exposure time is 30 s.
The nominal pixel scale and the full-width at half-maximum of the point spread function (PSF) are 1\arcsec .01 pixel$^{-1}$ and $\sim$ 2\arcsec, respectively \citep{ztf}.
The number of Aql X-1 data points are 157 and 640 for $\gp$ and $\rp$-band, respectively.

\subsection{LCO}\label{sec:lco}
Las Cumbres Observatory \citep{lco} is a world-wide network of optical telescopes
for time-domain astronomy.
We downloaded archived V and $\ip$-band reduced FITS images observed in 57460-58812 MJD
from LCO science archive\footnote{\url{https://archive.lco.global/}}.
The typical exposure time is 100 s.
The total number of images are 298 and 295 for V and $\ip$-band, respectively.
\autoref{tab:lco_images} summarized observatories and instruments contributed to the observations of LCO.
Low-quality images (e.g. tracking errors, blurred) were excluded before photometry.

\begin{table*}
    \caption{Observatories and instruments operated by LCO for Aql X-1 observations}
    \begin{tabular}{ccccc}
      \hline
      Observatory & Telescope$^*$ & Imager$^*$ & \multicolumn{2}{c}{Number of images}\\
       & & & V-band & $\ip$-band\\
      \hline
      Siding Spring Observatory, & 2m & Spectral & 123 & 122\\
      Australia & 1m & Sinistro & 1 & 1\\
      & 1m & SBIG STL-6303 & 1 & 1\\
      \hline
      Haleakala Observatory, Hawaii & 2m & Spectral & 152 & 150\\
      \hline
      South African Astronomical Observatory, & 1m & Sinistro & 3 & 3\\
      South Africa & 1m & SBIG STL-6303 & 4 & 5\\
      \hline
      McDonald Observatory, Texas & 1m & Sinistro & 1 & 1\\
      \hline
      Cerro Tololo Interamerican Observatory, & 1m & Sinistro & 6 & 5\\
      Chile & 1m & SBIG STL-6303 & 7 & 7\\
      \hline
      \multicolumn{5}{l}{
        $^*$ See \citet{lco} or LCO web-site (\url{https://lco.global/observatory/instruments/}).
      }
    \end{tabular}
    \label{tab:lco_images}
\end{table*}

\begin{figure}
    \includegraphics[width=\hsize]{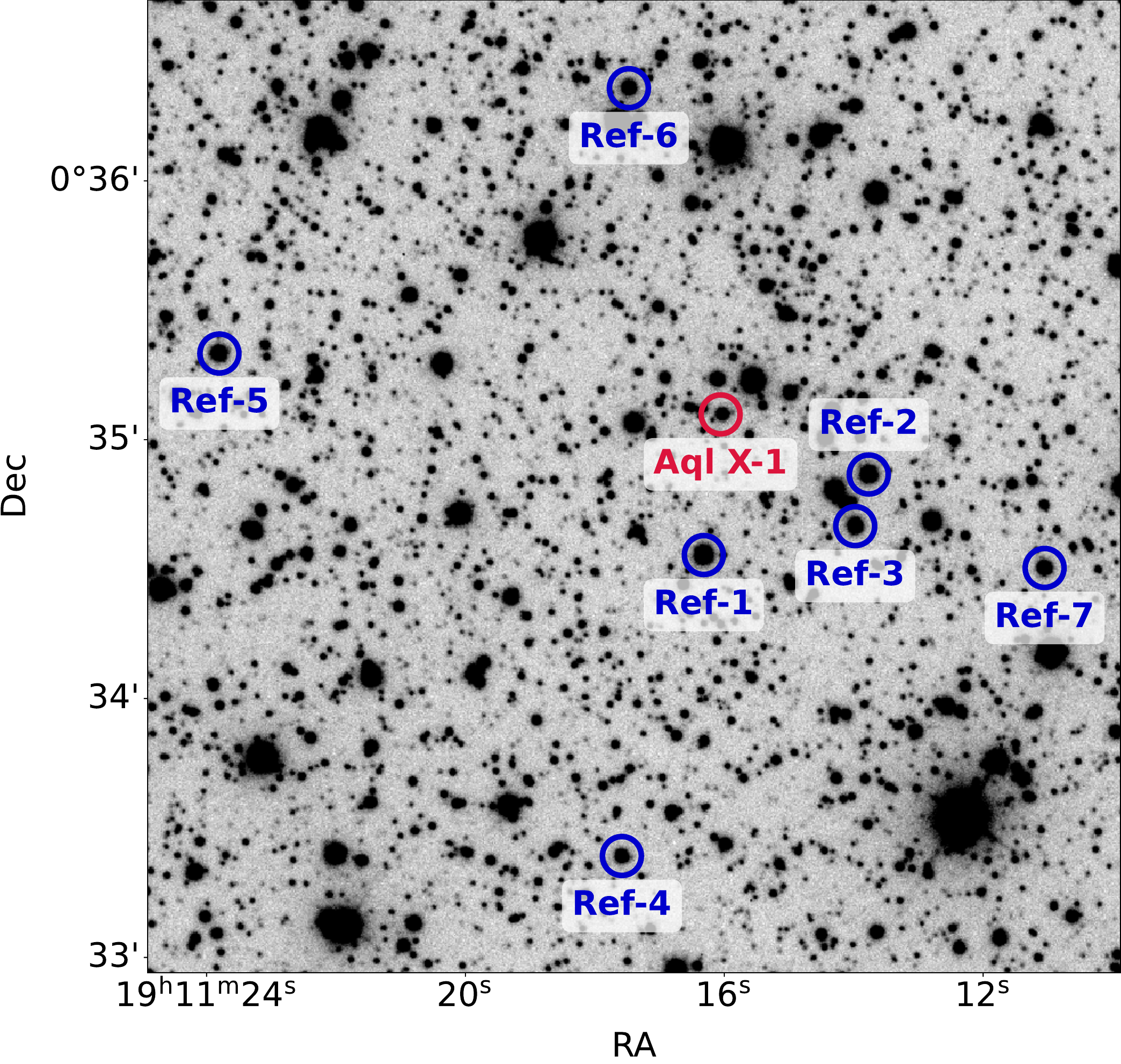}
    \caption{
        A finding chart of Aql X-1 field.
        Information of reference stars is described in \autoref{tab:ref-stars}.
        This is $\ip$-band image taken with 2m telescope at Haleakala Observatory operated by LCO.
    }
    \label{fig:finding-chart}
\end{figure}

We conducted photometry for LCO images by fitting the PSF using IRAF-\texttt{daophot} \citep{iraf}.
LCO telescopes cannot resolve the Aql X-1 and the interloper, and imaging them as almost a single point source.
However, the validity of the PSF fit considering them as a single source is questionable.
We therefore checked PSF subtracted images, and no systematic residuals were evident.
Presumably, the PSF distortion was negligible relative to the background noise because the separation of the two stars was smaller than the PSF size of LCO optics ($\sim$ 2\arcsec).
The magnitude estimation of Aql X-1 is performed by comparing with 7 local reference-stars, which are described in \autoref{tab:ref-stars}.
We obtained V and $\ip$-band magnitude of reference-stars from Swift/UVOT \citep{uvot} and Pan-STARRS \citep{ps1} catalog, respectively.
A finding-chart of Aql X-1 field is shown in \autoref{fig:finding-chart}.

\begin{table}
    \caption{Photometric reference stars for LCO images}
    \begin{tabular}{ccccc}\hline
      Label & RA & Dec & \multicolumn{2}{c}{Optical magnitude$^*$}\\
       & & & V-band & $\ip$-band \\
       & (H:M:S) & (D:M:S) & \multicolumn{2}{c}{(AB-magnitude)}\\\hline
      Ref-1 & 19:11:16 & 00:34:33 & 16.019(07) & 14.597(02) \\
      Ref-2 & 19:11:14 & 00:34:52 & 16.245(07) & 14.985(03) \\
      Ref-3 & 19:11:14 & 00:34:40 & 16.912(10) & 15.387(01) \\
      Ref-4 & 19:11:18 & 00:33:24 & 17.514(14) & 15.932(02) \\
      Ref-5 & 19:11:24 & 00:35:20 & 16.072(07) & 15.292(02) \\
      Ref-6 & 19:11:17 & 00:36:21 & 16.868(10) & 15.685(03) \\
      Ref-7 & 19:11:11 & 00:34:30 & 16.976(11) & 15.584(04) \\
      \hline
      \multicolumn{5}{l}{
        $^*$ Bracketed values are errors of last 2 digits.
      }
    \end{tabular}
    \label{tab:ref-stars}
\end{table}

\subsection{Correction of the dust extinction}\label{sec:reduction}
We estimated V-band extinction $A_\mathrm{V}=2.2 \pm 0.6$
from the observed hydrogen column density $N_\mathrm{H}=(4.4\pm0.1) \times 10^{21}\,\mathrm{cm}^{-2}$ \citep{nh} 
and the relation of $N_\mathrm{H}/A_\mathrm{V}=(2.0\pm0.5)\times10^{21}\,\mathrm{cm^{-2}mag^{-1}}$ \citep{nh2av}.
Extinctions of other bands were determined by using the relation of \citet{ccm}:
$A_\gp=2.6\pm0.7$, $A_\rp=1.9\pm0.5$ and $A_\ip=1.5\pm0.4$ for $\gp$, $\rp$ and $\ip$-band, respectively.

Optical magnitudes and fluxes shown hereafter are corrected the dust extinction unless otherwise noted.

\section{Result}\label{sec:result}

\subsection{Optical/X-ray light curves}\label{sec:lc}
\begin{figure*}
   \includegraphics[width=\hsize]{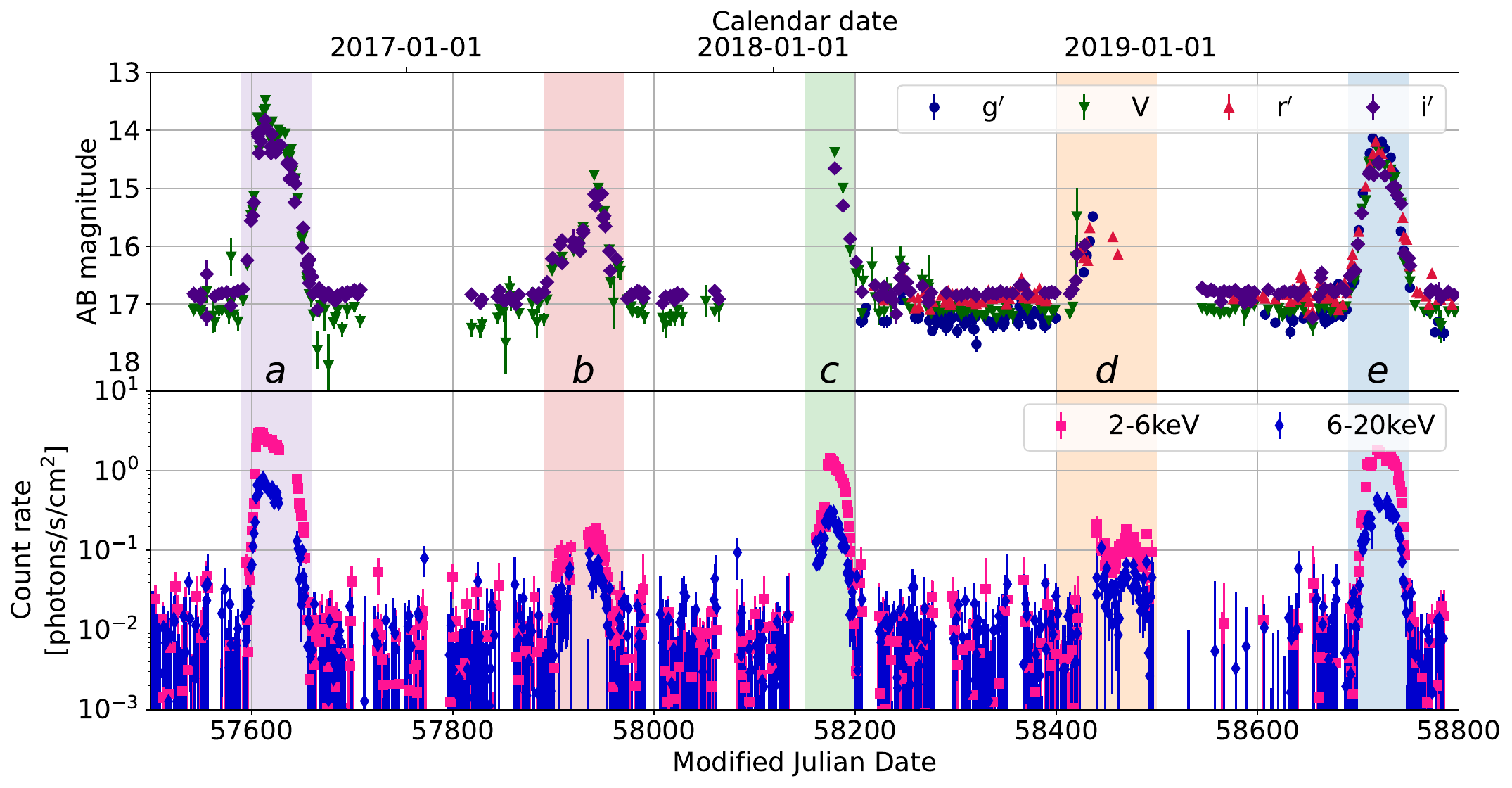}
   \caption{
        Optical and X-ray light curves of Aql X-1 in 57500-58800 MJD.
        The upper panel shows optical lightcurves and the lower shows X-rays.
        Outbursts with significant rise in both optical and X-ray bands occurred 5 times
        during this period.
        We labeled these 5 outbursts as outburst-$a$, $b$, $c$, $d$ and $e$.
        Colored areas are periods of each outburst, determined by the eyeball inspection.
        The optical 4-band data is available for only outburst-$d$ and $e$ because of ZTF observation epoch.
        Only a sparse optical data points is available around January due to the limitation of sun-angle.
        Note that optical magnitudes shown this figure are corrected the dust extinction (c.f. section~\ref{sec:reduction}).
    }
   \label{fig:lightcurve}
\end{figure*}

\autoref{fig:lightcurve} shows optical and X-ray light curves in 57500-58800 MJD.
Significant rise in both optical and X-ray light curves can be confirmed 5 times in this figure,
and those are consistent with the Astronomer's Telegram outburst reports \citep{atel_57600,atel_57900,atel_58150,atel_58400,atel_58700}.
We refers to outbursts occurred around 57600, 57900, 58150, 58400, and 58700 MJD
as outburst-$a$, $b$, $c$, $d$, and $e$, respectively.
In dark and stable periods (e.g. 58200-58400 MJD), the V-band magnitude is $\sim17$
(i.e. $\sim19$ without extinction correction) which is consistent with the V-band magnitude of the interloper \citep{chevalier}.
Therefore Aql X-1 was in a quiescent phase during these periods.
The optical 4-band data is available for only outburst-$d$ and $e$ because the ZTF data does not covers < 58194 MJD.
In addition, only a sparse optical data points are available for outburst-$c$ and $d$ due to the limitation of sun-angle.
Outburst-$a$, $c$, and $e$ are clearly brighter than $b$ and $d$ in both optical and X-ray band.
And near the peak, the former 3 outbursts have a large ratio of 2-6 keV count rates to 6-20 keV compared to the latter 2 outbursts.

It is difficult to study the quiescent phase of Aql X-1 with these optical data because the most optical photons came from the interloper.
Therefore we focused our study on the outburst period.
The difference in optical magnitude between the outburst and quiescent is 2-3 mag, and this means about 10\% of the optical flux during outburst should be a quiescent component.
We extracted the outburst flux by subtracting the quiescent flux.
The quiescent flux is determined by averaging the flux during 58300-58400 MJD, 
which is considered to be in the quiescent phase.
Obtained quiescent fluxes are as follows;
$0.45\pm0.06$, $0.52\pm0.03$, $0.65\pm0.05$ and $0.68\pm0.04$ mJy for $\gp$, V, $\rp$ and $\ip$-band, respectively.
On the other hand, X-ray fluxes in the quiescent period are only about 1\% of the peak flux,
which is less than errors during outbursts.
Therefore, the quiescent X-ray flux during outburst can be ignored.
From the next section, `flux' will mean the outburst flux unless otherwise noted.

\subsection{X-ray state transitions}\label{sec:xray}
\begin{figure}
    \includegraphics[width=\hsize]{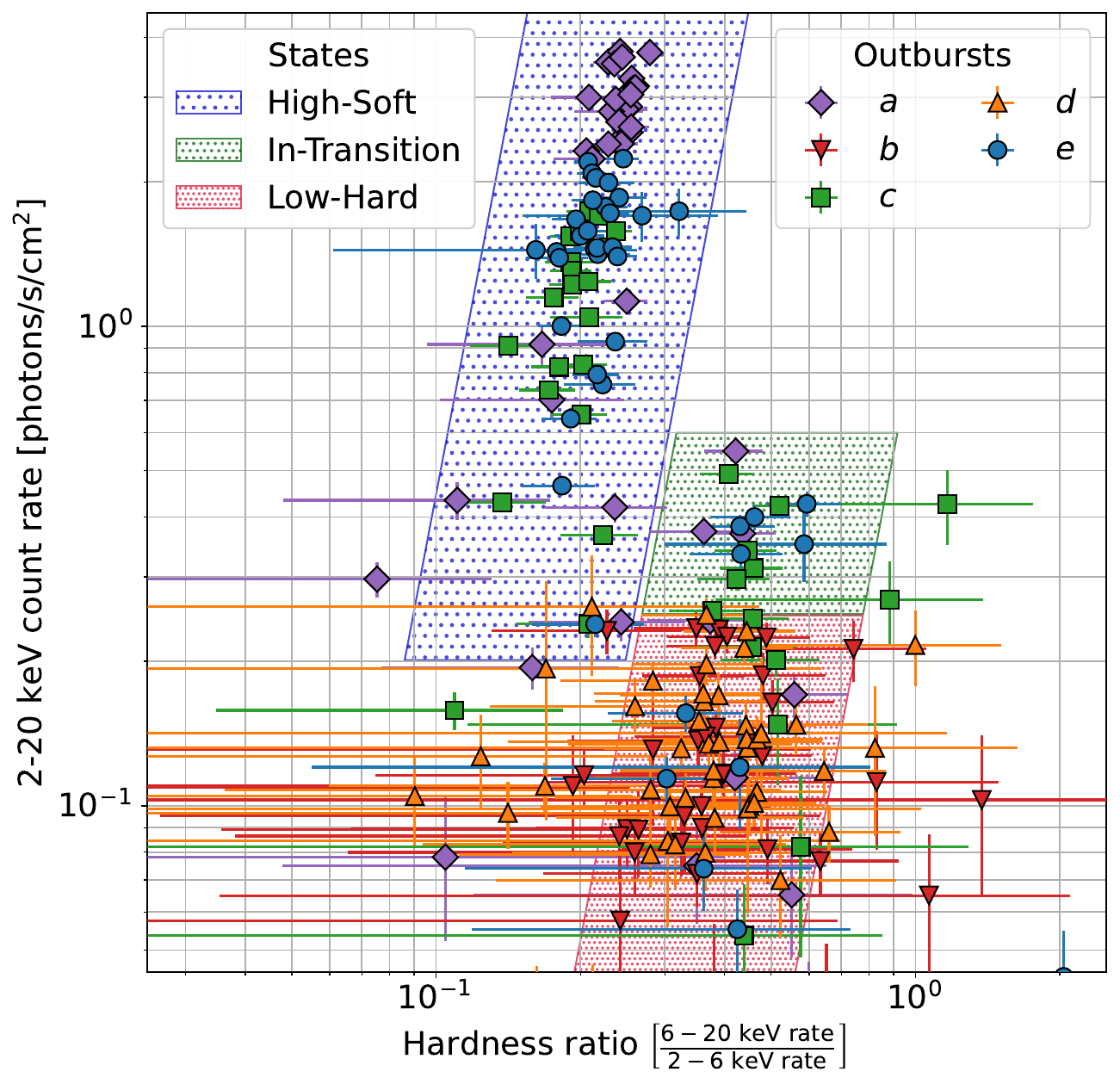}
    \caption{
        X-ray hardness-intensity diagram for 5 outbursts.
        Areas of LH, IT and HS state are determined by the eyeball inspection.
    }
    \label{fig:hid1}
\end{figure}

\begin{figure*}
    \includegraphics[width=\hsize]{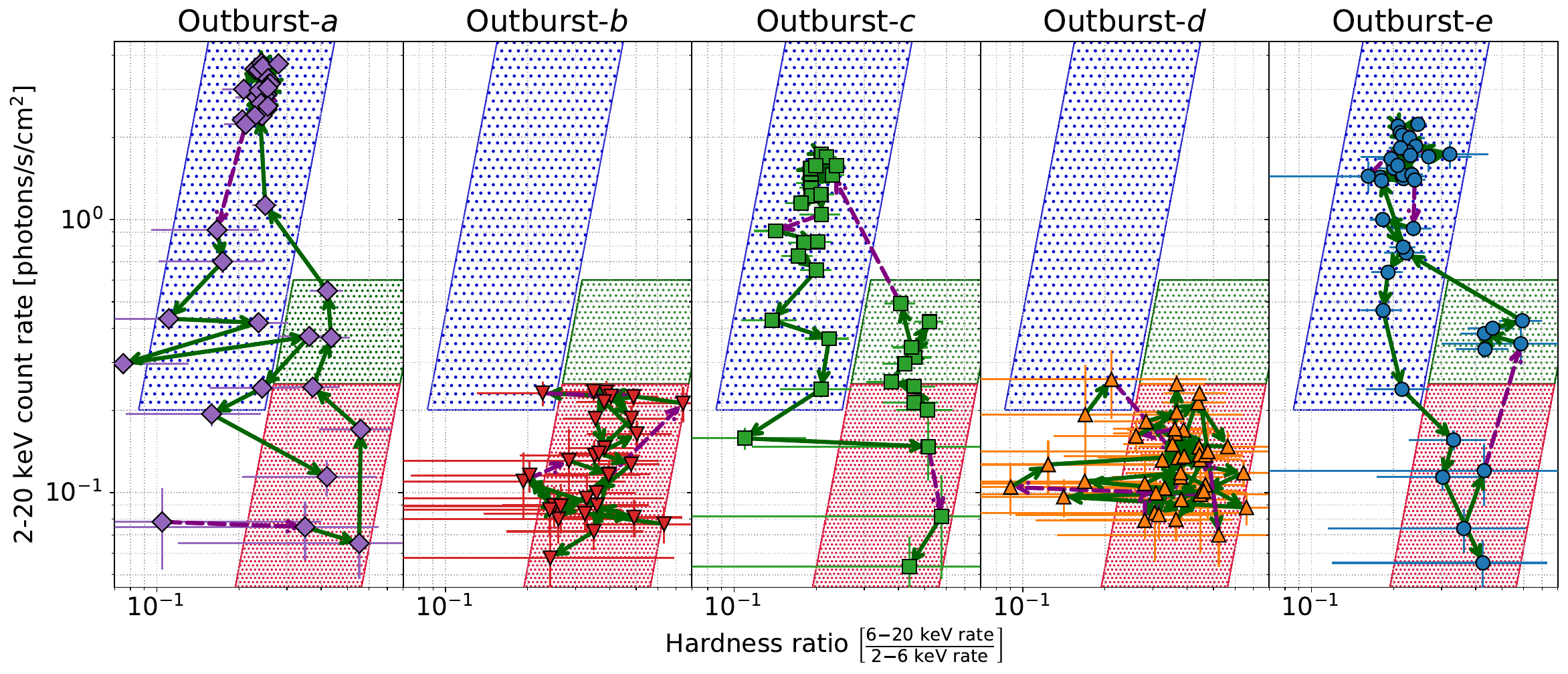}
    \caption{
        Movements of points on the hardness-intensity plane for 5 outbursts.
        The arrows connect each point to the next, and their style corresponding to a interval:
        green solid is $<$ 1.5 days and purple dashed is $\ge$ 1.5 days.
        The colored areas are the same as in \autoref{fig:hid1}.
    }
    \label{fig:hid2}
\end{figure*}

A X-ray hardness-intensity diagram (HID) is shown for the 5 outbursts in \autoref{fig:hid1}.
We defined a hardness ratio (HR) as (6-20 keV count-rate) / (2-6 keV count-rate).
Points are concentrated in hard and low-intensity area (HR $\gtrsim$ 0.3, intensity $\lesssim$ 0.6 $\cps$),
and soft and high-intensity area (HR $\lesssim$ 0.3, intensity $\gtrsim$ 0.6 $\cps$).
This indicates that Aql X-1 had at least two X-ray emission states in these outbursts, and we tentatively refer those states as `Low-Hard' (LH) and `High-Soft' (HS) respectively.
The outburst-$a$, $c$, and $e$ have points on both LH and HS state, but $b$ and $d$ have only on LH state.
Thus, the state transitions occurred in outburst-$a$, $c$, and $e$, but not in $b$ and $d$.
According to nomenclature of \citet{alabarta}, $a$, $c$, and $e$ are `full outbursts', and $b$ and $d$ are `failed-transition (FT) outbursts'.
They studied full and FT outbursts of BH-LMXBs, and found that no apparent difference between these 2 types of outbursts appears in the X-ray light curves, HID, and X-ray variability during the first tens of days after the outburst initiation.
Our data set is consistent with their work in that there is no difference in the distribution of full and FT outbursts on the HID in the same LH state.
\autoref{fig:hid2} shows movements of 5 outbursts on the HID.
In the case of full outbursts, it started in a dark and hard state, then brightened without much change in hardness, and transitioned to a soft state on a time scale of $\sim$ 1 day when the 2-20 keV intensity reached $\sim$ 0.5 $\cps$.
After the transition, it brightened further, then faded after passing the peak, and slowly transitioned to the hard state over a period of several days after the intensity drops $\lesssim$ 0.5 $\cps$.
On the other hand, unlike that full outbursts followed relatively simple paths, FT outbursts drew complex paths with repeating increases and decreases in intensity.
It should be noted that the intensity does not exceed $\sim$ 0.25 $\cps$ in FT outbursts.
This suggests that it can be in a transitional phase from LH to HS state when the intensity is between 0.25-0.6 $\cps$ in the brightening phase of the full outburst.
And we refer this as `In-Transition' (IT) state.

In this study, we determined the period between the 2-20 keV count rate exceeding 0.25 $\cps$ and exceeding 0.6 $\cps$ as IT, between exceeding 0.6 $\cps$ and falling below 0.2 $\cps$ as the HS, and the other as the LH.

The classification of LH and HS states is common in the X-ray binary study, but there is other terminology well summarized in \citet{hasinger}.
In this context, LMXBs are classified into `Z' and `atoll' sources based on their behavior on the X-ray HID and color-color diagram, and Aql X-1 belongs to the latter.
Then the state of atoll sources is divided into `banana branch' and `island branch', and they correspond to the HS and LH state, respectively in our terms.

\subsection{Optical/X-ray flux correlations}\label{sec:fred&lis}
\begin{figure}
    \includegraphics[width=\hsize]{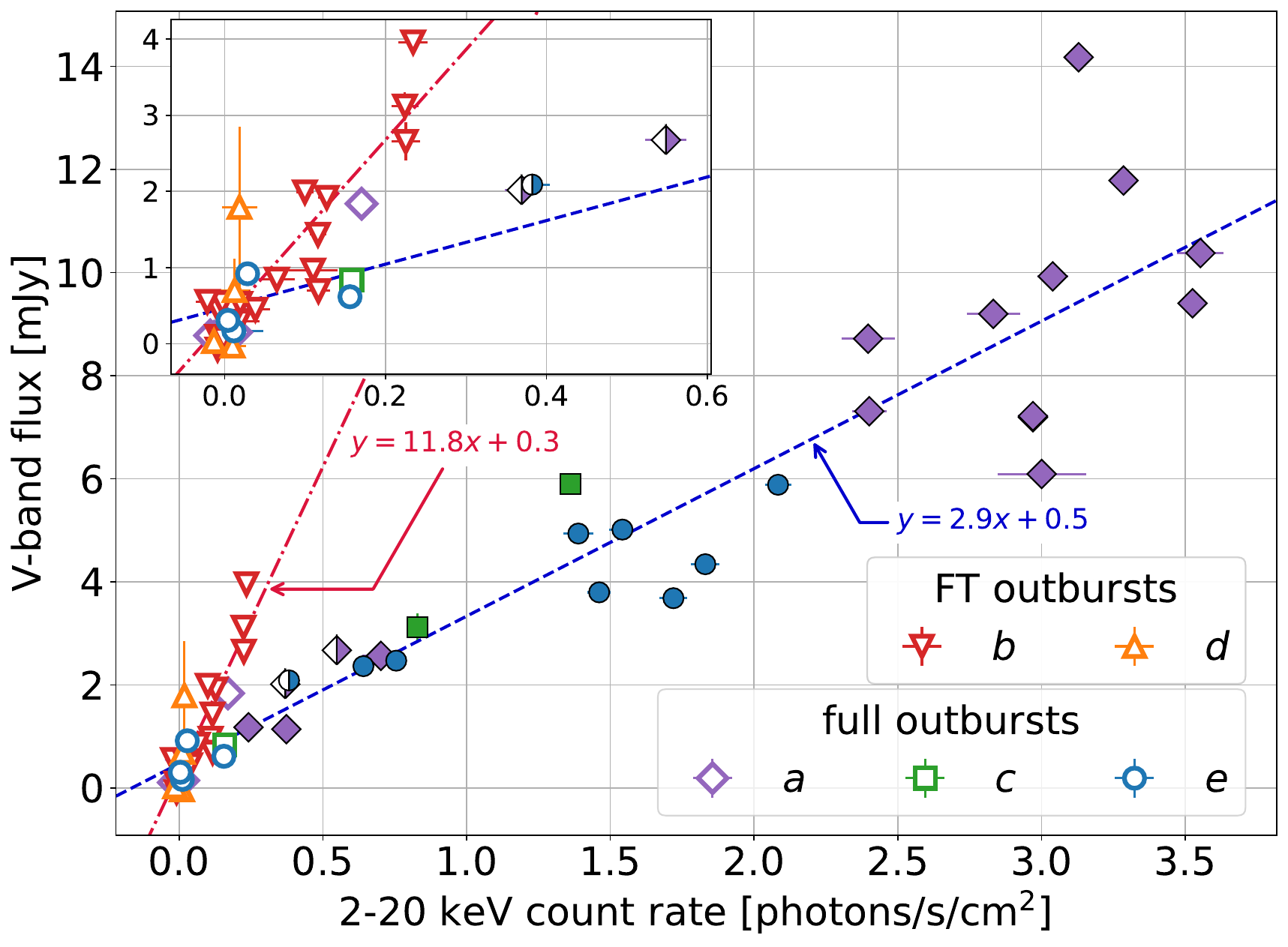}
    \caption{
        Scatter plots of 2-20 keV count rate and V-band flux for 5 outbursts.
        It is plotted with open markers in the LH state, half-filled markers in the IT state, and filled markers in the HS state.
        The shape and the color of the marker correspond to each outbursts.
        The blue dashed and red dotted-dashed lines are regression lines, fitted with HS and LH state respectively.
        The upper left panel is a partially enlarged view, where we plotted only points in the LH and IT state.
        Note that there were few simultaneous 2-20 keV and V-band data during outburst-$c$ and $d$.
    }
    \label{fig:ffp_v&x}
\end{figure}

\autoref{fig:ffp_v&x} shows scatter plots of V-band flux and 2-20 keV count rate for 5 outbursts.
We drew regression lines for each of LH and HS states, and their slopes differed by a factor of $\sim$ 4.
\citet{maitra_2008} classified the state of this source during outbursts into FRED and LIS according to the light curve shape, and found the difference in the optical/X-ray flux ratio between the two state.
In this context, the LH line would correspond to LIS and the HS line to FRED.
However, what these two lines indicate may not be the difference between LH and HS states, but between full and FT outbursts.
This is because the points of outburst-$c$ and $e$ whose 2-20 keV count rate $\sim$ 0.15 $\cps$ are closer to the line of the HS state, while they are in the LH state.
Although, it should be noted that these points are in the decay phase (i.e. after experiencing the state transition), and may be in the different state than the brightening phase, even in the same LH state.
In addition, the LH line is mostly determined by the points of the outburst-$b$, and this means that the difference in these lines may be due solely to the peculiarity of the outburst-$b$.
However, while there are only a few points on \autoref{fig:ffp_v&x}, the similarity of the optical/X-ray light curve shapes suggests that the behavior of outburst-$d$ on \autoref{fig:ffp_v&x} should be also similar to that of $b$.

\subsection{Optical CMD and SED}\label{sec:opt}
Since outburst-$a$, $b$, and $e$ have a sufficient V and $\ip$-band data, we performed the optical analysis for these 3 outbursts.

\begin{figure*}
   \includegraphics[width=\hsize]{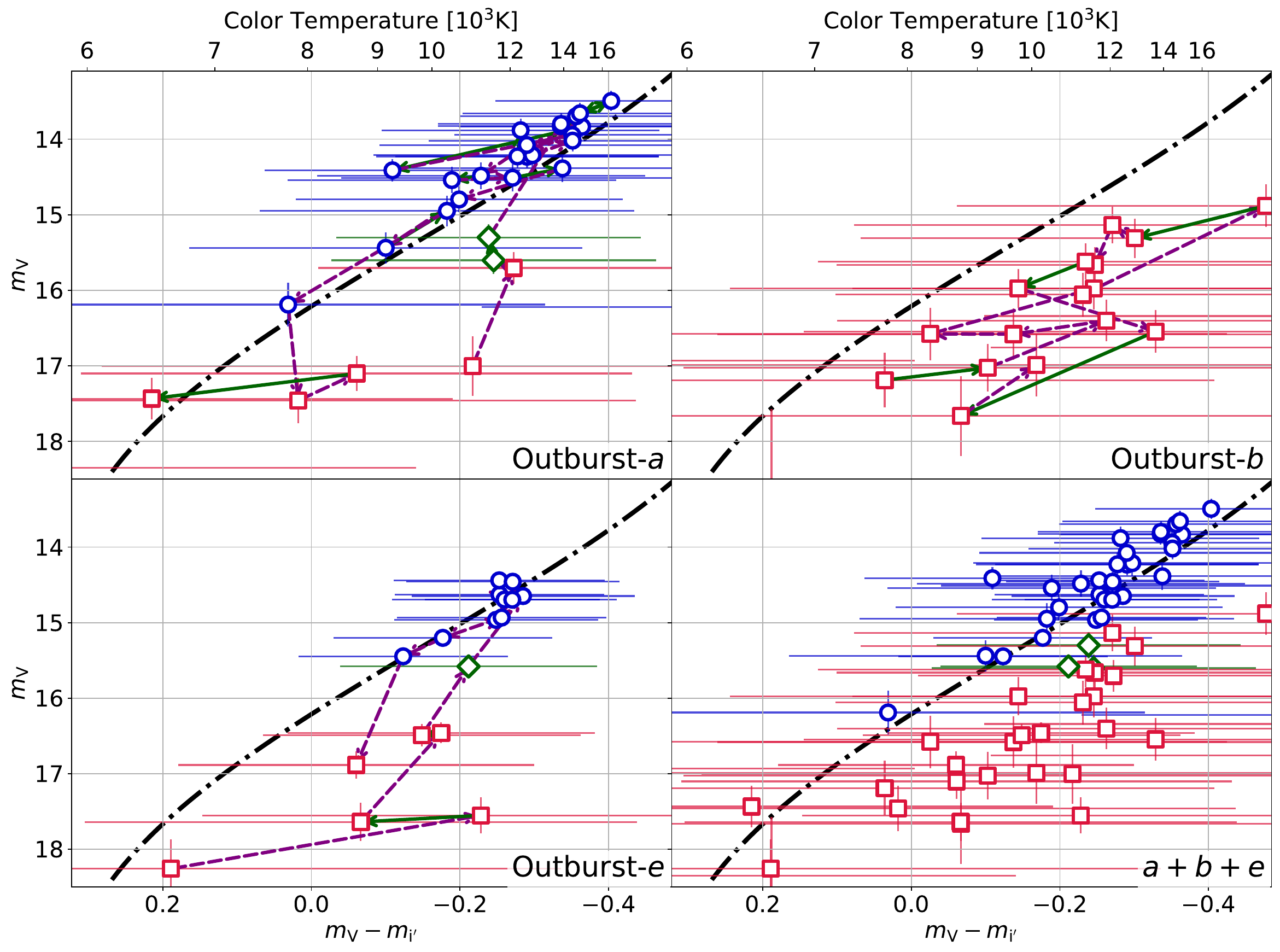}
   \caption{
        Color-magnitude diagrams for outburst-$a$, $b$ and $e$.
        $m_\mathrm{V}$ is the V-band magnitude,
        and $m_\mathrm{V}-m_\ip$ is the difference between the V and $\ip$-band magnitudes.
        These magnitudes are obtained by converting the outburst component of the flux (c.f. section~\ref{sec:lc}) with AB system.
        Note that the horizontal axis was set so that the right side is bluer, in accordance with the HID.
        The upper horizontal axis shows color temperature.
        The blue circle, the green diamond, and the red square marker correspond to HS, IT and LH state respectively.
        The arrows connect each point to the next,
        and their style corresponding to a interval: green solid is $<$ 1.5 days and purple dashed is $\ge$ 1.5 days.
        The dotted-dashed curve shows the color-magnitude variation of a simplified irradiated disk model with a varying irradiation power ($\Cirr\LX$) and a constant disk size ($\rout=1.5\times10^{11}$ cm).
        We plotted the 3 outbursts simultaneously in the bottom right panel.
    }
   \label{fig:cmd}
\end{figure*}

We showed color-magnitude diagrams (CMDs) for three outbursts in \autoref{fig:cmd}.
Magnitudes shown in this figure are obtained by converting the outburst component of the flux (c.f. section~\ref{sec:lc}).
Note that the horizontal axis was set so that the right side is bluer, in accordance with HIDs.
We also showed the color-magnitude variation of the simplified irradiated disk model (c.f. section~\ref{sec:discussion_opt}), in which an irradiation power (i.e. $\Cirr\LX$) varies and a disk size is constant ($\rout=1.5\times10^{11}$ cm).
The distance and inclination were set to 5 kpc \citep{li} and 42\degr \citep{sanchez}, respectively.
The full outbursts have a counterclockwise trajectory on these CMDs.
In other words, it brightened with blue state, slightly brightened while becoming blue after the state transition, and darkened while becoming red after the peak.
It is similar to the X-ray behavior on the HID in that the paths taken by the brightening and darkening are different and it is redder (softer) when darkening than the brightening (c.f. \autoref{fig:hid2}).
There is no significant difference in the distribution of the 3 outbursts on the color-magnitude plane for each HS, IT, and LH states.
In particular, both of the 2 full outbursts are correlated in color and magnitude in the HS state, consistent with the irradiated disk.
On the other hand, outburst-$b$ shows a color-magnitude correlation, albeit weaker than the HS state of full outbursts, and may be consistent with a smaller irradiated disk.
Although it is not clear due to a lack of data, the resemblance of the distribution on CMDs with outburst-$b$ suggests a similar color-magnitude correlation in the LH state of 2 full outbursts.

\begin{figure}
    \includegraphics[width=\hsize]{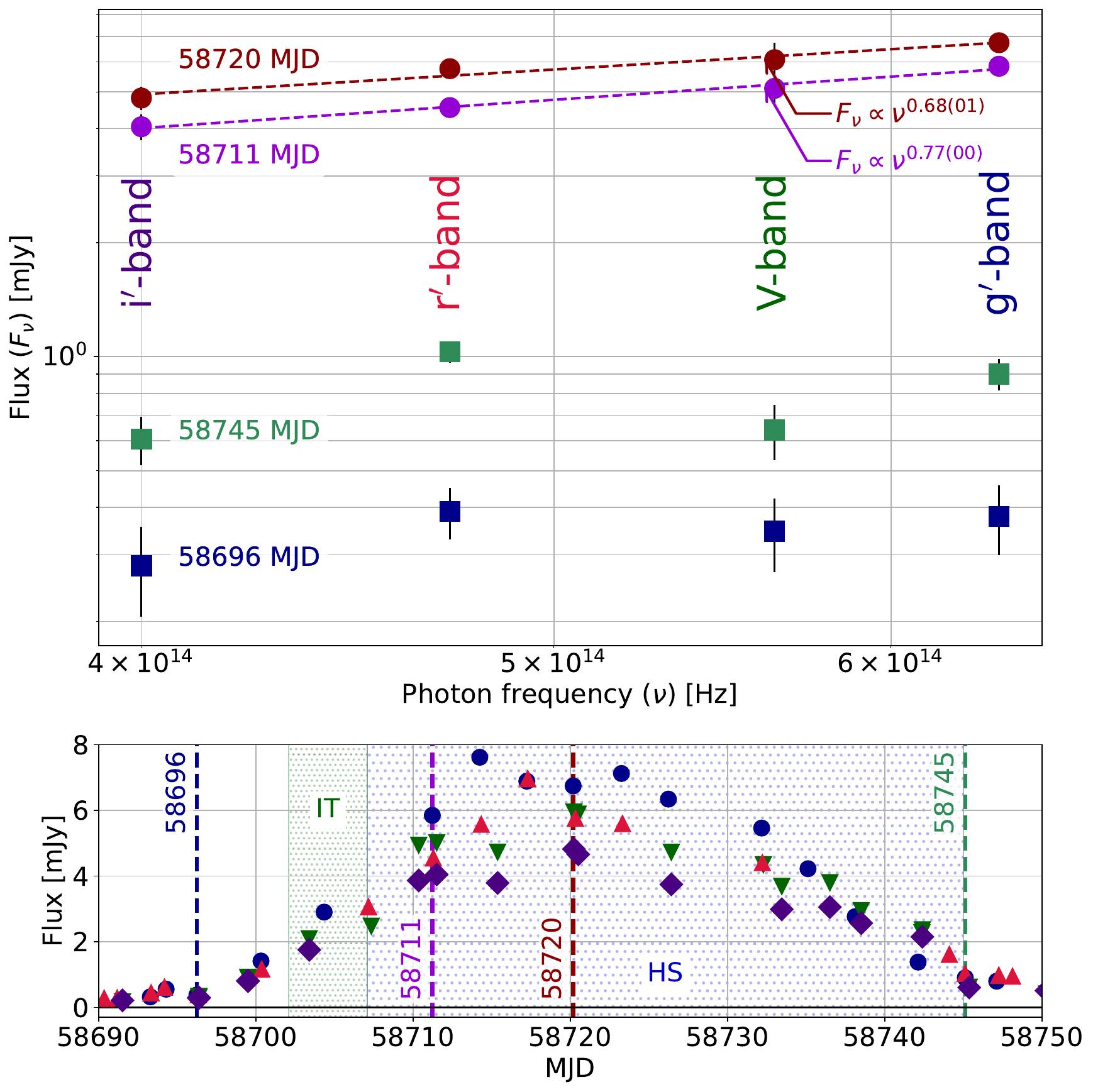}
    \caption{
        Optical spectral energy distribution and light curves of outburst-$e$ (top and bottom panel, respectively).
        (top) It is plotted for the 4 days when the simultaneous optical 4-bands data is available during the outburst.
        The marker color corresponds to the date,
        and the shape is circular in HS state and square in LH state.
        The dashed lines are fitted power-law spectra ($F_{\nu}\propto\nu^\alpha$) of the 2 days in the HS state.
        Note that this is a log-log plot.
        (bottom) The shape and color of the markers are the same as in \autoref{fig:lightcurve}.
        The vertical dashed lines correspond to the 4 dates, and their colors are the same as in top panel.
        Colored areas are the IT and HS period.
    }
    \label{fig:sed}
\end{figure}

During the outburst-$e$, there were 4 dates when simultaneous optical 4 band data available.
We showed optical spectral energy distribution (SED) for these dates in \autoref{fig:sed}.
In the HS state, it drew a straight line in the log-log plot, indicating the power-law like spectrum, and the spectral index $\alpha$ was about 0.7-0.8 ($F_{\nu}\propto\nu^{\alpha}$; $F_{\nu}$ is the flux, $\nu$ is the photon frequency).
In the LH state, it tends to be relatively bright in the ZTF data ($\gp$ and $\rp$-band) and dark in the LCO data (V and $\ip$-band).
We suspect that this is not a realistic spectral shape, but is affected by systematic errors due to differences in observational conditions (e.g. locations, instruments, and time) or photometric processes.

\subsection{Summary of results}\label{sec:summarize_result}
\begin{enumerate}
    \item Five outbursts ($a$-$e$) with significant rise in both optical and X-ray light curves were observed in 57500-58800 MJD.
    \item There are 3 X-ray emission state (LH, IT, and HS state) and the state transition occurred in outburst-$a$, $c$ and $e$, (full outbursts), while $b$ and $d$ remained in the LH state (FT outbursts).
    The LH and HS state correspond to the `island' and `banana' branch \citep{hasinger}.
    \item In the HS state, there was a optical color-magnitude correlation, which is consistent with a simplified irradiated disk model.
    \item The optical SED in outburst-$e$ was power-law like in the HS state with a spectral index $\alpha\sim0.7$ - $0.8$ ($F_{\nu}\propto\nu^{\alpha}$).
\end{enumerate}

\section{Discussion}\label{sec:discussion}
\subsection{Simplified irradiated disk model}\label{sec:discussion_opt}
Possible origins of optical radiation during outbursts of NS-LMXBs are multi-temperature blackbody radiation from the accretion disk and synchrotron radiation from the jet.
From section~\ref{sec:opt}, the optical emission in the HS state is consistent with the irradiated disk.
And the spectral index is $\alpha\sim0.7$-$0.8$ in optical band, which is difficult to explain with the jet synchrotron emission \citep{ryubicki}.
Therefore, the optical emission in the HS state is considered to be mainly multi-temperature blackbody radiation from the irradiated disk.

\begin{figure}
        \includegraphics[width=\hsize]{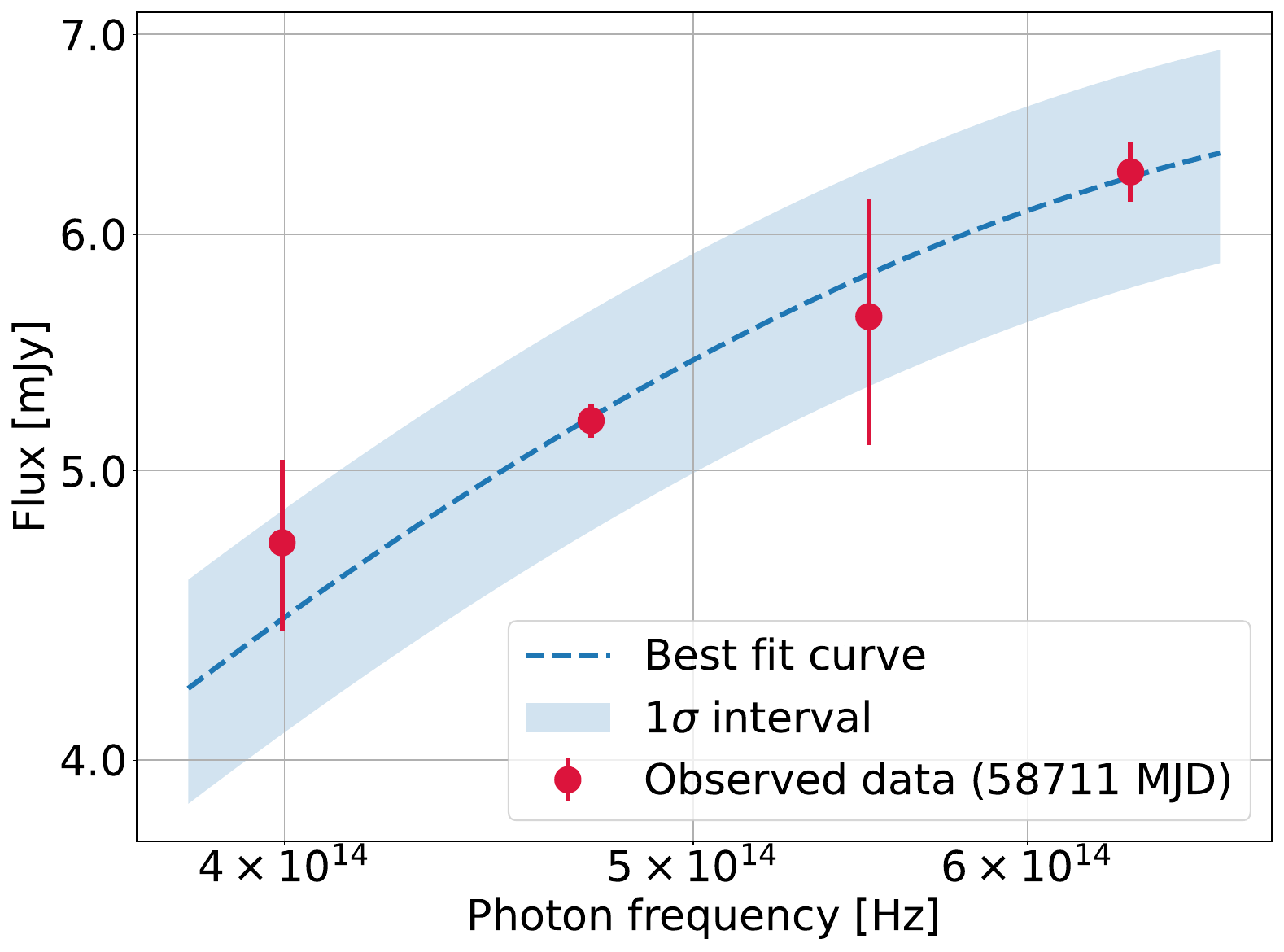}
        \caption{
            Fitted optical SED of irradiated disk model with observed values in 58711 MJD.
            Note that this is a log-log plot.
        }
    \label{fig:fitting}
\end{figure}

The irradiated disk is a model of accretion disk in which a geometrically thin and optically thick disk is heated by X-rays emitted from the vicinity of a compact object and emits thermal radiation \citep{paradijs,dubus1999,gierlinski}.
On the disk outer-rim, which is considered the main source of optical radiation, irradiation heating is expected to be more dominant than viscous heating in determining disk surface temperature.
Thus, we used a simplified irradiated disk model, which is constructed by ignoring viscous heating effect from the model of \citet{dubus2001}.
The model consists of following 2 formulae;
\begin{eqnarray}
    L_\nu &=& \int^{\rout}_{r_\mathrm{in}} B_\nu(T_r)\cdot2\pi r\diff{r}\label{eq:l_nu}\\
    \sigma T_r^4 &=& \frac{\Cirr\LX}{4\pi r^2}\label{eq:t_dist}
\end{eqnarray}
where $\nu$ is a photon frequency,
$L_\nu$ is a spectral luminosity at $\nu$,
$r$ is a radius from the neutron star,
$r_\mathrm{in}$ and $\rout$ are radii at the innermost and outermost edges of the disk respectively,
$B_\nu$ is a Planck distribution function,
$T_r$ is a disk surface temperature at $r$,
$\sigma$ is a Stefan–Boltzmann constant,
$\LX$ is an X-ray luminosity,
and $\Cirr$ is a fraction of the X-ray luminosity absorbed by the disk surface (hereafter `irradiation fraction').
$\Cirr$ contains information of the irradiation geometry, albedo, to name a few.
$\Cirr\LX$ represents the effective power of irradiation and is referred to as `irradiation power' in this paper.
For simplicity, we assumed that $\Cirr$ is uniform across the disk.
The distance and inclination were set to 5 kpc \citep{li} and 42\degr \citep{sanchez}, respectively.
We fitted the optical SED of 58711 MJD (near the peak of the outburst-$e$) using the \texttt{curve\_fit} function of \texttt{scipy.optimize} library \citep{scipy}, with fixing $r_\mathrm{in}$ to $9.8\times10^5$cm \citep{meshcheryakov}, and using $\rout$ and $\Cirr$ as free parameters.
$\LX$ was calculated as $3.8\times10^{37}$ erg/s using the 2-20 keV flux at this date ($8.6\times10^{-9}$ erg/s/cm$^2$).
The result is shown in \autoref{fig:fitting}, which shows that the model agrees the observed SED within 
$1\sigma$ interval.
Thus, the optical spectrum is consistent with the simplified irradiated disk model, at least in 58711 MJD.
Best fit parameters are $\Cirr=(3.4 \pm 0.3)\times10^{-3}$ and $\rout=(1.5 \pm 0.1)\times10^{11}$ cm.
This $\rout$ is comparable with the value obtained by \citet{meshcheryakov}.
On the other hand, $\Cirr$ is not simply comparable because they use an another parameter $f_\mathrm{out}$ (the product of $\Cirr$ and the ratio of $\LX$ to the accretion power) and performed fitting with fixing it.
In addition to the SED fit result, \autoref{fig:cmd} shows that optical color-magnitude variations in the HS state were consistent with the simplified irradiated disk of varying $\Cirr\LX$ and constant $\rout$.
Therefore, the optical emission during the HS state of Aql X-1 outbursts can be explained by the simplified irradiated disk also for periods other than 58711 MJD, and its variation is due to change in $\Cirr\LX$.

Incidentally, \autoref{fig:cmd} shows a color-magnitude correlation in the LH state which implies a presence of a irradiated disk smaller than in the HS.
There are three possible cases for the ``smaller irradiated disk''.
The first case is when the disk outermost radius is indeed smaller than that of the HS state.
In this case, the disk would have extended outward during the state transition by a certain mechanism such as viscous diffusion.
The second case is when the outer disk is not irradiated by X-rays.
For instance, the structure in the inner disk might impede X-rays from reaching the outer disk.
The last case is when the X-ray absorption efficiency of the outer disk is significantly lower.
In any case, it appears that the disk state transition propagated from the inside to the outside.

\subsection{Cause of variations in the irradiation}\label{sec:discussion_x}
\begin{figure}
    \includegraphics[width=\hsize]{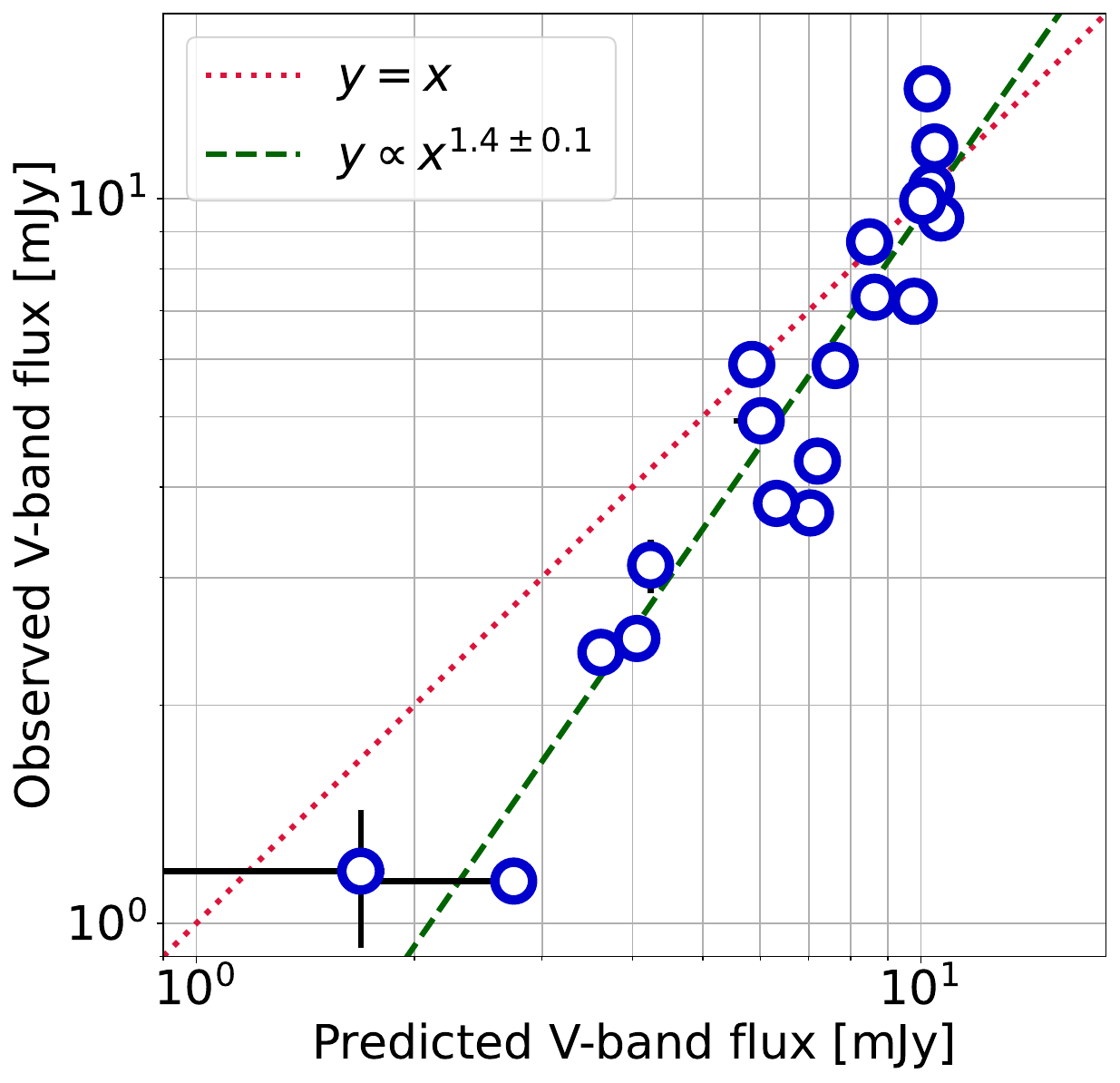}
    \caption{
        Predicted and observed V-band fluxes.
        The "predicted V-band flux" is derived from the observed 2-20 keV flux and the simplified irradiated disk model under the assumption of constant $\Cirr$ and $\rout$ (c.f. section~\ref{sec:discussion_x}).
        Errors of predicted fluxes are 90\% confidence interval.
        We plotted for epochs when there were simultaneous X-ray and V-band data in the HS state of 3 full outbursts.
        The red dotted line represents the scenario where the predicted and the observed flux are equal, and the green dashed line is the regression line.
    }
    \label{fig:x-l_acc}
\end{figure}

The contribution of $\Cirr$ and $\LX$ to the irradiation cannot be separated by only analyzing optical variations.
However, the fact that the X-ray flux varied in correlation with optical-band (c.f. \autoref{fig:lightcurve}, \ref{fig:ffp_v&x}) shows that at least $\LX$ varied and should be contributed to optical variations.
Thus, we verified whether the optical and X-ray variations are consistent under the assumption of constant $\Cirr$ by comparing the observed and the model-predicted V-band flux.
We derived the predicted V-band flux $F_\mathrm{V, pred}$ by following equation;
\begin{eqnarray}
    F_\mathrm{V, pred} = \frac{\int F_{\nu}\tau_{\nu}\diff{\nu}}{\int \tau_{\nu}\diff{\nu}}\label{eq:f_prediction}
\end{eqnarray}
where $\nu$ is a photon frequency, $F_{\nu}$ is a flux density distribution of the model, and $\tau_{\nu}$ is a transmittance at $\nu$ of the V-band filter.
$F_{\nu}$ is calculated from the 2-20 keV flux with fixed parameters of $\Cirr=3.4\times10^{-3}$ and $\rout=1.5\times10^{11}$ cm.
We downloaded the transmittance table of the V-band filter from the LCO official web-page\footnote{\url{https://lco.global/observatory/instruments/filters/}}, and derived $\tau_{\nu}$ by an unit conversion and a linear interpolation.
The result is shown in \autoref{fig:x-l_acc}.
The observed flux was proportional to the 1.4 power of the predicted flux.
This means that the observed variability of the V-band flux was larger than that predicted from the 
observed X-ray flux and the constant $\Cirr$.

There are two possible causes for this result.
The first is an underestimation of the $L_X$ variability.
We calculated $\LX$ from the 2-20 keV flux, but this may underestimate the $\LX$ variability because the X-ray spectrum tends to be `harder when brighter' in the HS state (c.f. \autoref{fig:hid1}).
Thus we obtained the 20-30\footnote{The nominal upper-limit of the GSC energy band is 30 keV \citep{maxi}.} keV flux using the same procedure as for the 2-20 keV flux (c.f. section~\ref{sec:maxi}), and it was only $\sim0.1\%$ of the 2-20 keV flux.
Since > 30 keV X-rays are assumed to be even less, we determined that the contribution of > 20 keV X-rays to the $\LX$ is negligible.
However, not all factors which could cause underestimation have been eliminated.
The second is a positive correlation between $\Cirr$ and $\LX$.
In this case, there should exist a certain mechanism which make the disk to absorb X-rays more efficiently in conjunction with the X-ray brightening.
The $\Cirr$ is proportional to the solid angle occupied by a unit area of the disk surface in perspective of the X-ray source.
When the outer disk is irradiated, it is qualitatively expected that the heated matter on the disk expands and the disk thickness increases.
Then $\Cirr$ is thought to increase as the solid angle of the outer disk increases.
Therefore $\Cirr$ and $\LX$ are anticipated to have a positive association.

\begin{figure}
    \includegraphics[width=\hsize]{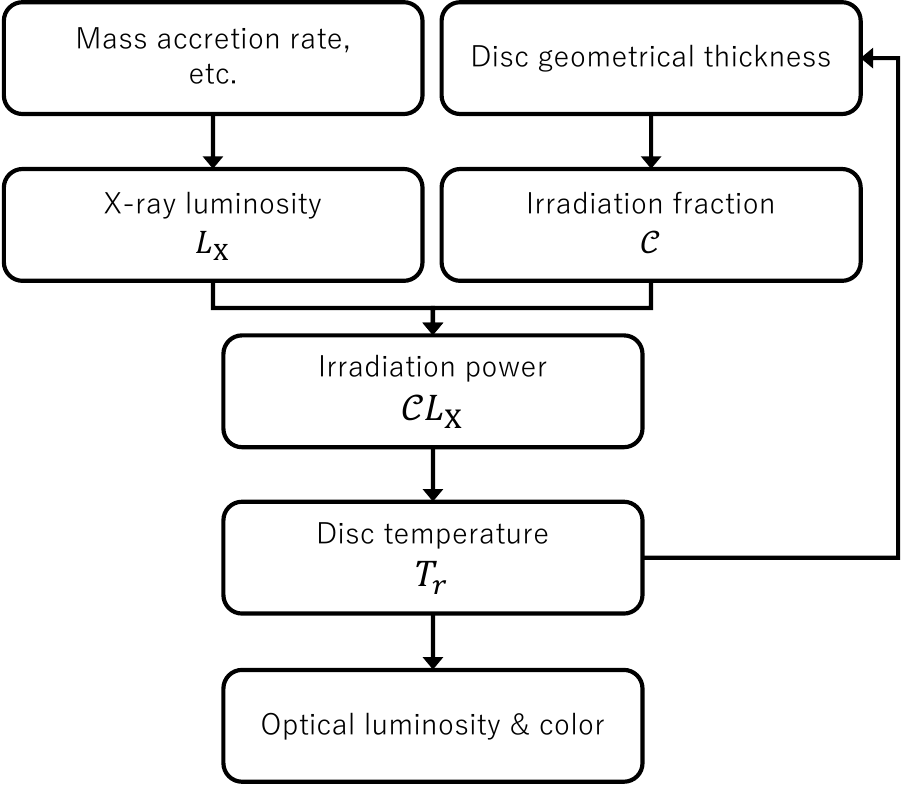}
    \caption{
        Possible scenario for the optical/X-ray variation.
        The boxes mean quantities, and the arrows indicate causality between variations of quantities which has positive correlation.
        (For example, the disk temperature $T_r$ rises as irradiation power $\Cirr\LX$ increases.)
    }
    \label{fig:scenario}
\end{figure}

Our suggested scenario is illustrated as a flowchart in \autoref{fig:scenario}.
In brightening phase, the X-ray luminosity $\LX$ is increased by some factor (e.g. mass accretion rate).
Since the disk temperature $T_r$ is coupled with $\LX$ by the irradiation, the optical luminosity also increases and the color becomes bluer.
At the same time, the rise in $T_r$ causes the disk thickness to expand, which increases the irradiation fraction $\Cirr$.
This process results in a positive feedback on irradiation power $\Cirr\LX$, and the optical brightening becomes greater than that estimated from the X-ray brightening.
The opposite variation occurs in the same order during the decay phase.

It should be noted that our explanation about the $\Cirr$ variation due to the disk expansion is only qualitative.
There must be various factors involved in determining the disk thickness other than gas pressure, and the temperature rise alone may not be adequate to cause sufficient expansion.
In addition, we did not discuss about the variation in the disk surface albedo, which probably depends on the irradiating X-ray spectrum and the disk surface temperature.

\section{Summary}\label{sec:summary}
We studied optical and X-ray behaviors of the LMXB Aql X-1 using X-ray data of MAXI and optical data of ZTF and LCO in 57500-58800 MJD.
As a result, we confirmed 5 optical/X-ray outbursts in this period, and X-ray state transitions across three states (LH, IT, and HS) occurred in 3 outbursts (full outbursts) while remain 2 outbursts stayed in the LH state (FT outbursts).
In the HS state, optical color/magnitude correlation and optical SED were consistent with the simplified irradiated disk model.
The observed optical variability was larger than that predicted from the X-ray variability, but could be qualitatively explained by considering the positive feedback to the irradiation power caused by the variation of the disk thickness.

\section*{Acknowledgements}

We would like to express our sincere gratitude to the anonymous reviewers for their valuable comments and suggestions on the interpretation of the optical CMDs.
Additionally, we would like to extend our appreciation for their helpful recommendations of relevant literature.
M. Niwano was supported by JSPS KAKENHI Grant-in-Aid for JSPS Research Fellow.
This work was partially supported by JSPS KAKENHI Grant-in-Aid for Scientific Research 19H00698 and JP23H01214.
This research utilized MAXI data provided by RIKEN, JAXA, and the MAXI team.
We acknowledge the Samuel Oschin 48-inch Telescope at the Palomar Observatory as part of the Zwicky Transient Facility project, supported by the National Science Foundation under Grant No. AST-1440341 and a collaboration including Caltech, IPAC, the Weizmann Institute for Science, the Oskar Klein Center at Stockholm University, the University of Maryland, the University of Washington, Deutsches Elektronen-Synchrotron and Humboldt University, Los Alamos National Laboratories, the TANGO Consortium of Taiwan, the University of Wisconsin at Milwaukee, and Lawrence Berkeley National Laboratories. Operations are conducted by COO, IPAC, and UW.
We also acknowledge the Las Cumbres Observatory global telescope network for providing observations.
The Pan-STARRS1 Surveys (PS1) and the PS1 public science archive have been made possible through contributions by the Institute for Astronomy, the University of Hawaii, the Pan-STARRS Project Office, the Max-Planck Society and its participating institutes, the Max Planck Institute for Astronomy, Heidelberg, and the Max Planck Institute for Extraterrestrial Physics, Garching, The Johns Hopkins University, Durham University, the University of Edinburgh, the Queen's University Belfast, the Harvard-Smithsonian Center for Astrophysics, the Las Cumbres Observatory Global Telescope Network Incorporated, the National Central University of Taiwan, the Space Telescope Science Institute, the National Aeronautics and Space Administration under Grant No. NNX08AR22G issued through the Planetary Science Division of the NASA Science Mission Directorate, the National Science Foundation Grant No. AST-1238877, the University of Maryland, Eotvos Lorand University (ELTE), the Los Alamos National Laboratory, and the Gordon and Betty Moore Foundation.
We acknowledge the Astronomy Data Center (ADC), National Astronomical Observatory of Japan, for the use of the Multi-wavelength Data Analysis System.
Finally, we thank the Astropy community \footnote{http://www.astropy.org} for developing the core Python package for Astronomy \citep{astropy1, astropy2}, which was utilized for data analysis in this research.

\section*{Data Availability Statement}
All of the optical/X-ray data underlying this article are publicly available.
MAXI/GSC data we used was obtained from MAXI on-demand web interface (\url{http://maxi.riken.jp/mxondem/}).
ZTF light curves are available in NASA/IPAC infrared science archive (\url{https://irsa.ipac.caltech.edu/Missions/ztf.html}).
LCO reduced images are available in LCO science archive (\url{https://archive.lco.global/}).



\bibliographystyle{mnras}
\bibliography{main}



\appendix


\bsp	
\label{lastpage}
\end{document}